\documentclass[%
 reprint,
superscriptaddress,
 amsmath,amssymb,
 aps,
pra,
]{revtex4-1}
\usepackage{graphicx}
\usepackage{subfigure}
\usepackage{braket, amsfonts}
\usepackage{amssymb}
\usepackage{color}
\usepackage{hyperref}
\usepackage{mathbbol}
\usepackage{bm}

\begin{document}


\title{Green function formalism for resonant interaction of x-rays
 with nuclei in structured media}


\author{Xiangjin Kong}
\email{xjkong@mpi-hd.mpg.de  }
\affiliation{Max-Planck-Institut f\"ur Kernphysik, Saupfercheckweg 1, D-69117 Heidelberg, Germany}

\affiliation{Department of Physics, National University of Defense
	Technology, Changsha, China}

\author{Darrick E. Chang}
\affiliation{ICFO-Institut de Ciencies Fotoniques, The Barcelona Institute of Science and Technology, 08860 Castelldefels (Barcelona), Spain}
\affiliation{ICREA-Instituci\'o Catalana de Recerca i  Estudis Avan\c{c}ats, 08015 Barcelona, Spain}

\author{Adriana P\'{a}lffy}
\email{palffy@mpi-hd.mpg.de}
\affiliation{Max-Planck-Institut f\"ur Kernphysik, Saupfercheckweg 1, D-69117 Heidelberg, Germany}


\date{\today}


\begin{abstract}

The resonant interaction between x-ray photons and nuclei is one of the most exciting subjects of the burgeoning field of x-ray quantum optics. A resourceful platform used so far are thin-film x-ray cavities with embedded layers or M\"ossbauer nuclei such as $^{57}\mathrm{Fe}$. A new quantum optical model based on the classical electromagnetic Green's function is developed to investigate theoretically the nuclear response inside the x-ray cavity. The model is versatile and provides an intuitive picture about the influence of the cavity structure on the resulting spectra. We test its predictive powers with the help of the semiclassical coherent scattering formalism simulations  and discuss our results for increasing complexity of layer structures.

\end{abstract}




\maketitle


\section{Introduction}

Compared to optical photons, x-rays have a number of desirable properties such as 
deeper penetration, better focus, no longer limited by an inconvenient diffraction limit as for low-frequency photons, and correspondingly superior spatial resolution, as well as robustness, and the large momentum transfer they may produce. The commissioning of the first X-ray Free Electron Lasers (XFELs) \cite{LCLS-web, SACLA-web} has brought into attention all the advantages of x-ray photons and supports the development of the emerging field of x-ray quantum optics \cite{Rohringer2012,Adams2013}. However, the modest degree of control that we have over x-ray photons is at present a major drawback. 

X-ray quantum optics with nuclei promises to close this gap by exploiting the resonant interaction of x-rays with M\"ossbauer nuclear transitions. For instance, by using the 14.4 keV nuclear resonance in $^{57}\mathrm{Fe}$, methods have been proposed  and  experimentally implemented to coherently control single x-ray photons. A number of experimental achievements have rendered x-ray quantum optics a burgeoning field, among which the  storage of nuclear excitation via magnetic switching \cite{Shvydko1996},
the observation of the collective Lamb shift  \cite{rohlsberger2010collective}, electromagnetically induced transparency with x-rays \cite{rohlsberger2012electromagnetically}, the first experimental evidence of vacuum-generated coherences \cite{PhysRevLett.111.073601}, slow x-ray light \cite{heeg2015tunable}, the manipulation of single-photon wavepacket pulse shapes \cite{vagizov2014coherent},  interferometric phase detection at x-ray energies  \cite{heeg2015interferometric}, the collective strong coupling of single x-ray photons \cite{haber2016,haber2017rabi} and the controlled spectral narrowing of x-ray pulses \cite{heeg2017spectral}.  These developments provide potential applications for the fields of metrology, material science, quantum information, biology and chemistry.  From the theoretical side, several works have addressed promising control schemes for stopping and manipulating x-ray quanta
\cite{palffy4,palffy1,kong2014field,kong2016stopping,gunst2016logical,Liao2018,zhang2019nuclear}.

\begin{figure}
	\includegraphics[width=0.5\textwidth]{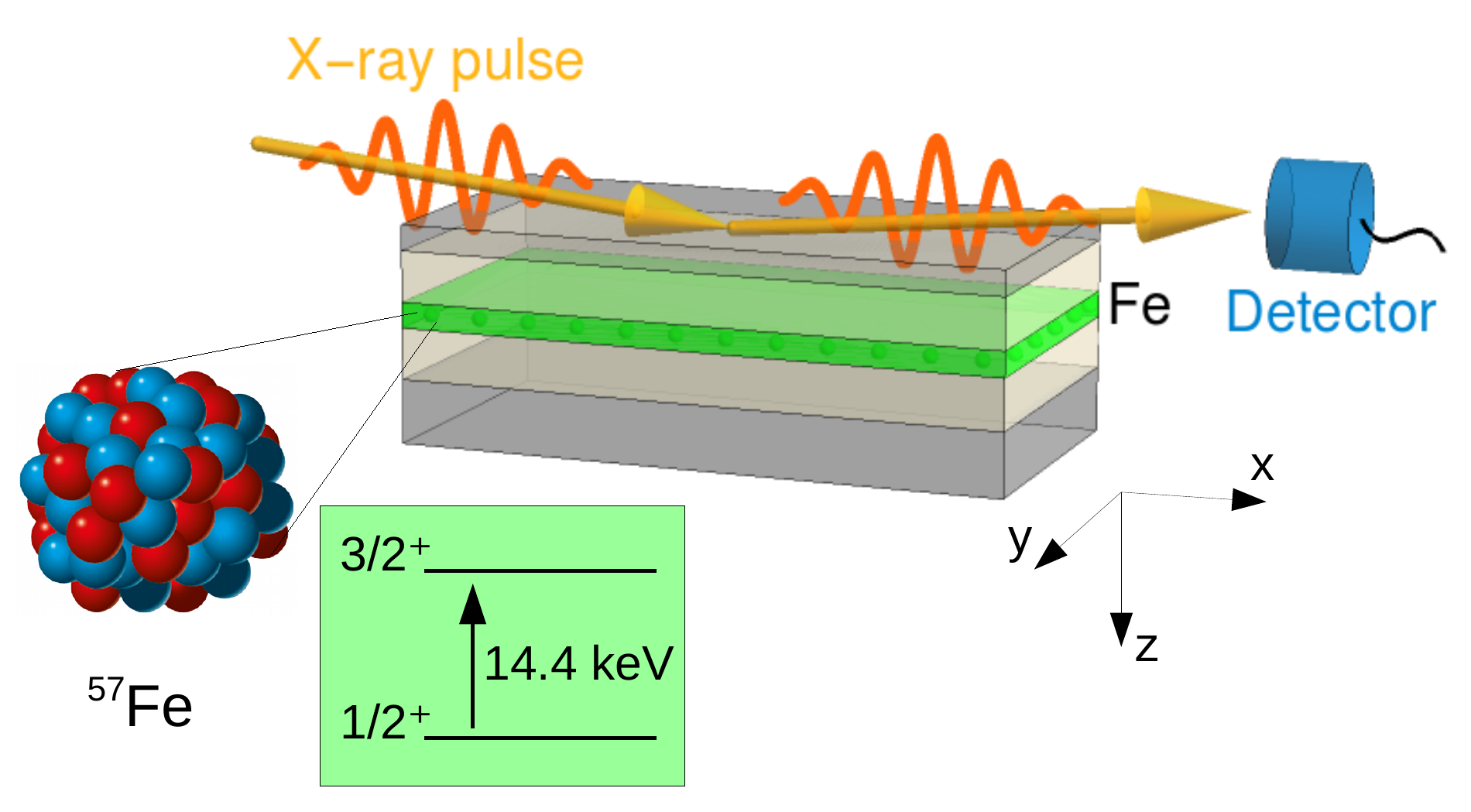}
	\caption{\label{system} Sketch of the x-ray thin-film cavity with a nuclear ensemble containing $^{57}\mathrm{Fe}$ (green layer). For the top and bottom layers usually a high-$Z$ material such as Pd or Pt are used. The low-density guiding layers can be fabricated for instance from C or B$_4$C. }
\end{figure}

Very successful physical systems for x-ray quantum optics applications are thin-film x-ray cavities - see an illustration in Fig.~\ref{system}. A thin-film x-ray cavity  typically comprises of a stack of stratified materials. A low-density guiding layer is coated on a substrate with higher electron density (high atomic number $Z$) in a planar geometry. The incident x-rays, typically produced at synchrotron radiation facilities, arrive at grazing angle and couple evanescently to the cavity forming a standing wave. The nuclear layer is placed in the cavity and it interacts with this standing wave allowing a better control over the resonant interaction. Many of the experimental achievements of x-ray quantum optics were based on x-ray thin-film cavities \cite{rohlsberger2010collective,rohlsberger2012electromagnetically,PhysRevLett.111.073601,heeg2015interferometric,heeg2015tunable,haber2016,haber2017rabi}.

In the past slightly more than fifty years several methods have been developed to describe theoretically the x-ray quantum dynamics in crystals or thin-film cavities. Shortly after the discovery of the M\"ossbauer effect, at the end of the 1960s, a quantum theory for x-ray and $\gamma$-ray optics for crystals containing resonant nuclei was established by Hannon and Trammell \cite{hannon1968mossbauer,hannon1969mossbauer,hannon1974mossbauer} using quantum electrodynamics $S$-matrix techniques. Neglecting the possibility of coherent multi-photon effects (which were at the time not to be anticipated while working with M\"ossbauer sources) the theory produced in the weak-excitation limit a set of coupled equations of the multiple scattering type formally identical to  those derived in the dynamical x-ray theory in a semiclassical manner \cite{Authier2001}. These expressions were particularized also for grazing incidence \cite{hannon1969mossbauer}, and followed by a row of studies on grazing incidence antireflection films for synchrotron radiation, among which also pure nuclear reflections were investigated \cite{Hannon1985}. For normal and Bragg incidence, further important theoretical developments and detailed treatments of the dynamical theory in the semiclassical limit (treating the scattered field classically) were given by Afanas'ev, Kagan, and co-workers (see for a review Ref.~\cite{Kagan1999} and references therein) or by Shvyd'ko \cite{Shvydko1999}. 

More concretely for the case of thin-film cavities, semi-classical methods such as  the Parratt formalism \cite{PhysRev.95.359} or the layer formalism \cite{rohlsberger2004coherent} implemented in the software package CONUSS \cite{PhysRevB.49.9285,sturhahn2000conuss} have proven themselves very successful in modeling  experimental data \cite{rohlsberger2010collective,rohlsberger2012electromagnetically,PhysRevLett.111.073601,heeg2015interferometric,heeg2015tunable,haber2016,haber2017rabi}. This is remarkable considering that  so far the low intensity of synchrotron radiation sources allows mostly single resonant x-ray photons to couple to the M\"ossbauer nuclei in the cavity.  Due to the classical nature of the x-ray field in these methods, it is however  impossible  to study the quantum properties of the x-ray photons, for instance, to calculate higher order correlation functions. This point might become important in experiments with XFEL light, where each pulse can contain more than a single resonant photon. A first XFEL experiment in nuclear forward scattering geometry for thick samples has been already performed \cite{Chumakov2018}, with up to 68 resonant photons per pulse. An even higher photon degeneracy could be reached with seeded XFELs or at an XFEL Oscillator \cite{XFELO}. We note that the quantum description dating back to the original works of Hannon and Trammell is also restricted to single photons \cite{hannon1968mossbauer,hannon1969mossbauer}.
A second draw-back concerns the difficulty to predict the structure starting from the desired scattering properties. Both the Parratt formalism or the computer package CONUSS can successfully predict the scattering spectra starting from structure, but cannot be easily used for the inverse problem.

A phenomenological quantum model for the x-ray cavities has been developed several years ago \cite{PhysRevA.88.043828,heeg2015collective} and used to model experimental data for specific cases. While quite versatile for single-layer cavities \cite{PhysRevA.88.043828}, the original  model had difficulties to accurately describe more complicated structures, and an extension including  multiple modes in the cavity was required to correctly reproduce experimental data for cavities with two embedded nuclear layers \cite{heeg2015collective}.  Both models can handle an arbitrary number of excitations. While Refs.~\cite{PhysRevA.88.043828,heeg2015collective} have focused on the regime of single excitations, a situation which corresponds to the so-far studied case of synchrotron radiation driving the nuclear transitions, the case of  stronger excitation up to population inversion was discussed in Ref.~\cite{Heeg-arXiv2016}.  A just recently developed general ab-initio few-mode model for quantum potential scattering problems  promises to be applicable also for x-ray thin-film cavities \cite{PhysRevX.10.011008}.  We note that after the submission of this work, another ab initio approach using Green's functions for nuclear quantum optics in x-ray cavities was brought to our attention \cite{Lentrodt2020}.

In this work, we develop a different formalism for the scattering of x-ray radiation off thin-film cavities, taking into account the nucleus-nucleus interaction in terms of the classical electromagnetic Green's function \cite{asenjo2017atom,UlrichD}.  The approach that we develop here describes the atom-light interactions using a quantization scheme based on the classical electromagnetic Green's function. The classical propagator describes the wave propagation between two atoms (in our case nuclei), while the quantumness of the system is encoded in the correlations of the local polarization noise operators and in the atoms (nuclei) as quantum sources \cite{asenjo2017atom}.  As such the field is treated quantum-mechanically and quantum observables are accessible although the field propagation obeys the wave equation and  the spatial profile of the photons is determined by the classical propagator. In our special geometry, the cavity structure determines the strength of the inter-nuclear coupling. The thin-film cavity is treated as a quasi-1D nanostructure and the cavity fields are effectively eliminated.  

Our formalism is very general and convenient to apply for complicated multi-layer structures. The model is not restricted to single excitations and therefore useful for future applications involving XFEL light. As it accounts for the quantization of  the   field, the model  
can be used to investigate the quantum properties of x-ray photons, for instance via higher order correlation functions.  These features are shared with the previously existing quantum models developed in Refs.~\cite{PhysRevA.88.043828,heeg2015collective}. We benchmark the semiclassical observables of our model by using CONUSS \cite{PhysRevB.49.9285,sturhahn2000conuss} to simulate spectra for several layer structures with one, two or thirty embedded $^{57}\mathrm{Fe}$ layers, the latter being the first attempt to  quantitatively describe with a quantum model a complex structure investigated experimentally in Ref.~\cite{haber2016}. The comparison shows perfect agreement between the two methods and confirms the validity of our formalism. We use the model also to predict and discuss the shape of the superradiant decay and the electromagnetically induced x-ray transparency results from Ref.~\cite{rohlsberger2012electromagnetically}.

 We note that within the field of quantum optics with neutral atoms, Green's function based approaches are also being actively explored. This formalism has recently been used to predict and quantitatively model several exotic, strongly correlated quantum optical phenomena, which cannot clearly be obtained by other means. Examples include the demonstration of photon number sorting via strong photon number-dependent group velocities \cite{manzoni2017simulating,sah2019dynamics}, the ``fermionization'' of excitations \cite{PhysRevX.7.031024,PhysRevA.99.023802} and the emergence of critical many-body dynamics \cite{PhysRevA.99.023802}. Alongside the development of better sources and optimized devices, we anticipate that the Green's function approach could facilitate the arrival of strongly quantum phenomena in x-ray optics.

This paper is organized as follows. In Sec.~\ref{2}, we introduce the effective Hamiltonian, in which the cavity fields are eliminated and the nucleus-nucleus interaction is written in terms of the classical Green's function \cite{asenjo2017atom}. We then write down the expressions for some observables using the input-out formalism \cite{asenjo2017atom}.  Next,  we present and analyze numerical results for three layer structures with one, two and thirty embedded $^{57}\mathrm{Fe}$ layers in Sec.~\ref{numres}. The results are compared to semiclassical simulations and several physical systems investigated experimentally are discussed in the light of the new model.   Finally we summarize our findings and point out  possible applications in Sec.~\ref{4}.

\section{Theoretical model}
\label{2}
In this Section we present the model formalism starting from a more detailed description of the physical system, the model Hamiltionian and master equation that rules the dynamics and the input-output formalism determining the observables in experiments. 

\subsection{Thin-film cavities}

Specific layer geometries allow x-rays to be guided in thin-film cavities. The thin-film consists of nm-thick layers of different materials, with an example  depicted in Fig.~\ref{system}.  A material of low electron density (carbon or boron carbide) is sandwiched between two layers of high electron density (palladium or platinum, for instance). Depending on the film thickness, a certain number of guided modes can be excited at different incidence angles where the reflectivity reaches a deep minimum. These minima determine the so-called resonant angles for the cavity and their positions depend on the cavity structure. Usually in experiments one measures reflectivity spectra for angles close to such a resonant angle and defines the deviation angle $\Delta\varphi=\varphi-\varphi_C$ where $\varphi$ is the incidence angle and $\varphi_C$ is the constant resonant angle \cite{PhysRevA.88.043828}. 

In order to have the x-rays drive a nuclear M\"ossbauer resonance, a $^{57}\mathrm{Fe}$ layer is embedded in  the thin-film cavity typically within the lossy low-$Z$ material layer. This isotope has a stable ground state and a first excited state at 14.413 keV, corresponding to a wavelength of 0.86 $\AA$. These two states are connected via a magnetic dipole ($M1$) transition.  With the nuclear resonance of approx.~4.66 neV natural linewidth, even when tuned to the nuclear transition energy, both synchrotron and XFEL pulses will act as broadband sources. In the case of the synchrotron,   just one and very rarely two resonant photons are available in each pulse at best. In the following we will develop a model that considers the interaction between the nuclei and the total electric field in the presence of a dispersive and absorptive medium which is spatially inhomogeneous.

\subsection{Model Hamiltonian and inter-nuclear couplings}

 The general Hamiltonian based on the quantum optics approaches in Refs.~\cite{PhysRevA.46.4306,PhysRevA.56.5022,PhysRevA.57.3931,Suttorp_2004,PhysRevLett.122.213901} in the rotating-wave approximation contains atomic, field and interaction terms,
\begin{align}
\hat{ H}=& \hat{H}_{atom}+\hat{H}_{field}+\hat{H}_{int} \nonumber\\
 =&\hbar\omega_0\sum_{i=1}^{N}\hat{\sigma}_{eg}^{i}\hat{\sigma}_{ge}^{i}+\hbar\int d\mathbf{r}\int_{0}^{\infty}d\omega\omega\mathbf{\hat{f}}^\dagger\left(\mathbf{r},\omega\right)\cdot\mathbf{\hat{f}}\left(\mathbf{r},\omega\right) \nonumber \\
 &-\left(\sum_{i=1}^{N}\hat{\sigma}_{eg}^{i}\int_{0}^{\infty}d\omega\,\mathbf{d}_i\cdot\mathbf{\hat{E}}\left(\mathbf{r}_i,\omega\right)+\text{H.c.} \right)
 \label{ho}
\end{align}
%
where $\hbar$ is the reduced Planck constant, $\omega_{0}$  the transition frequency,  $\hat{\sigma}_{eg}^{i}=| e \rangle \langle g |$ and  $\hat{\sigma}_{ge}^{i}=| g \rangle \langle e |$  denote the atomic raising and lowering operators of the $i$th atom, respectively,  and $\mathbf{d}_i$  its  dipole moment matrix element \cite{asenjo2017atom}.  Without loss of generality, we assume here two-level atoms such that all have the same dipole moment matrix element denoted by $\mathbf{d}$. The sum over $i$ runs over all $N$ atoms (nuclei) interacting with the photon field. The latter is described by the  bosonic annihilation and creation operators $\mathbf{\hat{f}}\left(\mathbf{r},\omega\right)$ and $\mathbf{\hat{f}}^\dagger\left(\mathbf{r},\omega\right)$, respectively, which satisfy the canonical commutation relations \cite{PhysRevA.57.3931,asenjo2017atom}.

The electric field operator $\mathbf{\hat{E}}\left(\mathbf{r},\omega\right)$
 fulfills the equation \cite{PhysRevA.57.3931,PhysRevA.53.1818,PhysRevLett.122.213901}
\begin{equation}
\bigtriangledown\times\bigtriangledown\times\mathbf{\hat{E}}\left(\mathbf{r},\omega\right)-\frac{\omega^2}{c^2}\varepsilon(\mathbf{r},\omega)\mathbf{\hat{E}}\left(\mathbf{r},\omega\right)=i\omega\mu_{0}\mathbf{\hat{j}}_{\text{noise}}(\mathbf{r},\omega)\, ,
\label{e1}
\end{equation}
where $\varepsilon(\mathbf{r,\omega})$ is the complex permittivity function describing the medium, $c=1/\sqrt{\mu_{0}\varepsilon_0}$ is the speed of light in vacuum,  $\varepsilon_0$ is the permittivity of free space, $\mu_0$ is the vacuum permeability and $\mathbf{\hat{j}}_{\text{noise}}(\mathbf{r},\omega)=\omega\sqrt{\left(\hbar\varepsilon_0/\pi\right)\text{Im}\left[\varepsilon(\mathbf{r},\omega)\right]}\mathbf{\hat{f}}(\mathbf{r},\omega)$.
 A formal solution of Eq.~(\ref{e1}) can be derived using the system's Green function $\mathbf{G}\left(\mathbf{r},\mathbf{r}^\prime,\omega\right)$ which satisfies \cite{PhysRevA.57.3931,PhysRevA.53.1818,asenjo2017atom}
\begin{equation}
\bigtriangledown\times\bigtriangledown\times\mathbf{G}\left(\mathbf{r},\mathbf{r}^\prime,\omega\right)-\frac{\omega^2}{c^2}\varepsilon(\mathbf{r},\omega)\mathbf{G}\left(\mathbf{r},\mathbf{r}^\prime,\omega\right)=\delta\left(\mathbf{r}-\mathbf{r}^\prime\right)\mathbf{I}\, ,
\end{equation}
where $\mathbf{I}$ is the unity dyadic. The electric field operator at frequency $\omega$ can be written in terms of the Green function and the annihilation (creation) field operators $\mathbf{\hat{f}}$\, ($\mathbf{\hat{f}^\dagger}$) as \cite{PhysRevA.57.3931,PhysRevA.53.1818,asenjo2017atom}
\begin{align}
	\mathbf{\hat{E}}\left(\mathbf{r},\omega\right)=&i\mu_{0}\omega^2\sqrt{\frac{\hbar\varepsilon_0}{\pi}}
\nonumber \\	
	&\times \int d\mathbf{r}^\prime\sqrt{\text{Im}\left[\varepsilon(\mathbf{r},\omega)\right]}\mathbf{G}\left(\mathbf{r},\mathbf{r}^\prime,\omega\right) \cdot\mathbf{\hat{f}}(\mathbf{r}^\prime,\omega)\, .
\end{align}
The total field operator is then
\begin{equation}
 \mathbf{\hat{E}}(\mathbf{r})=\int d\omega\mathbf{\hat{E}}\left(\mathbf{r},\omega\right)+\text{H.c.}\, 
\end{equation}

The expressions written so far apply to a broad class of problems. We now particularize this approach to our problem of interest. First, we note that nuclear transitions resonant to x-rays are often not of electric dipole type, and in particular the  $^{57}\mathrm{Fe}$ M\"ossbauer transition has magnetic dipole multipolarity. This translates to the use of the reduced nuclear transition probability $B(M1)$ \cite{Ring1980} instead of the electric dipole operator for matrix elements of the Hamiltonian (\ref{ho}). The exact expression for interaction Hamiltonians going beyond the dipole approximation in nuclear quantum optics can be found in Refs.~\cite{Palffy2008,kong2014field}. For simplicity and in order to keep the parallel to atomic quantum optics, we continue to use in the following the electric dipole moment matrix element $\mathbf{d}$ in our expressions.
Second, 
 we want to study the evolution of $N$ identical nuclei which interact via the probe x-ray field in the thin-film cavity.  The single-nucleus coupling strength to the cavity remains much smaller than the cavity linewidth.  This allows us to use the Born-Markov approximation and trace out the photonic degrees of freedom \cite{asenjo2017atom,PhysRevA.66.063810,Chang_2012}. The dynamics of the system can be described by means of the master equation \cite{Scully2006}
%
\begin{equation}
	\dot{\hat{\rho}}=-\frac{i}{\hbar}\left[\hat{H},\hat{\rho}\right]+L[\hat{\rho}]\, ,
	\label{master}
\end{equation}
 where $\hat{\rho}$ is the density matrix of the system and $L[\hat{\rho}]$ is the Lindblad operator modeling its loss. The resulting effective Hamiltonian  is written now explicitly in terms of the nucleus-nucleus interaction,
\begin{align}
	\hat{H}=&-\hbar\Delta\sum_{i=1}^{N}\hat{\sigma}_{eg}^{i}\hat{\sigma}_{ge}^{i}-\hbar\sum_{i,j=1}^{N}g^{ij}\hat{\sigma}_{eg}^{i}\hat{\sigma}_{ge}^{j}  \nonumber \\
	&-\sum_{i=1}^{N}\left[\mathbf{d}\cdot\hat{\mathbf{E}}_{p}^- (\mathbf{r}_{i})\hat{\sigma}_{ge}^{i}+\mathbf{d}^*\cdot\hat{\mathbf{E}}_{p}^{+}(\mathbf{r}_{i})\hat{\sigma}_{eg}^{i}\right]\, ,
	\label{h1}
\end{align}
where $\hat{\mathbf{E}}_{p}$ is the probe field and the notations ${\mathbf{E}}^{+(-)}$ were introduced for the positive (negative) frequency components of the field operator \cite{asenjo2017atom}. Furthermore,   $\Delta=\omega_p-\omega_0$ is the detuning between the  probe field $\omega_p$ and the nuclear transition with energy $\hbar\omega_0$.   For the thin-film geometry which possesses translational symmetry in the  $(x,y)$ plane, the probe field can be written as $\hat{\mathbf{E}}^{\pm}_{p}=\hat{\mathbf{E}}^{\pm}_{\text{1D}}(z)e^{\pm i\bm{k}^{\rho}_{ p}\bm{\rho}} $, where $\bm{\rho}=(x,y)$ and  $\bm{k}^{\rho}_p=((k_{p})_x,(k_{p})_y)$ is the transversal component of the incident wave vector $\bm{k}_p$.
The loss in the system is described by  the  Lindblad operator  \cite{asenjo2017atom}
\begin{align}
	L[\hat{\rho}]=-\sum_{i,j=1}^{N}\frac{\gamma^{ij}}{2}\left(\hat{\sigma}_{eg}^{i}\hat{\sigma}_{ge}^{j}\hat{\rho}+\hat{\rho}\hat{\sigma}_{eg}^{i}\hat{\sigma}_{ge}^{j}-2\hat{\sigma}_{ge}^{i}\hat{\rho}\hat{\sigma}_{eg}^{j}\right)\, .
	\label{l1}
\end{align}
In Eqs.~(\ref{h1}) and (\ref{l1}), we have introduced the spin-exchange and decay rates defined as 
\begin{align}
	g^{ij}&=\left(\mu_{0}\omega_{p}^{2}/\hbar\right)\mathbf{d}^*\cdot \text{Re}\left[\mathbf{G}\left(\mathbf{r}_{i},\mathbf{r}_{j},\omega_p\right)\right]\cdot \mathbf{d}\, ,  \nonumber \\
	\gamma^{ij}&=\left(2\mu_{0}\omega_{p}^{2}/\hbar\right)\mathbf{d}^*\cdot \text{Im}\left[\mathbf{G}\left(\mathbf{r}_{i},\mathbf{r}_{j},\omega_p\right)\right]\cdot \mathbf{d}\, . 
	\label{c1}
\end{align}
Note that the nucleus-nucleus couplings are given in terms of the total Green's function of the medium. Thus, if the Green's function is calculated either numerically or analytically, then the spin-exchange and decay rates and in turn the effective Hamiltonian and the Lindblad operators can be obtained. This will allow us to study the dynamics and properties of the system and the scattered photons using the master equation (\ref{master}).

\subsection{Green function for thin-film cavity geometry}

The Green function for the thin-film layers geometry has been derived analytically in Ref.~\cite{tomavs1995green}.  Exploiting the translational invariance of the system in the $(x,y)$ plane, the thin-film x-ray cavity is treated as a quasi-1D structure along the $z$ direction. The  complex permittivity function $\varepsilon(\mathbf{r},\omega)=\varepsilon(z,\omega)$ is defined in a stepwise fashion, according to the geometry illustrated in Fig.~\ref{system}.  The Green function can be written   as \cite{tomavs1995green}
\begin{eqnarray}
 &&\mathbf{G}(\mathbf{r}_{i},\mathbf{r}_{j},\omega_p)= \nonumber \\ 
 && \frac{1}{(2\pi)^2}\int d^2\bm{k}^{\rho}\mathbf{G}_{\text{1D}}(z_i,z_j,\omega_p,\bm{k}^{\rho})e^{i\bm{k}^{\rho}(\bm{\rho}_i-\bm{\rho}_j)}\, .
\label{G2}
\end{eqnarray}
The quantity $\mathbf{G}_{\text{1D}}(z_i,z_j,\omega_p,\bm{k}^{\rho})$ is a one-dimensional Green function  for the $z$ direction and differs in dimension from $\mathbf{G}(\mathbf{r}_{i},\mathbf{r}_{j},\omega_p)$ by an area factor. Ref.~\cite{tomavs1995green} provides the expression of $\mathbf{G}_{\text{1D}}(z_i,z_j,\omega_p,\bm{k}^{\rho})$ for multilayers, which can be simplified for a small incidence angle $\varphi\ll 1$, weak polarization dependence and $\bm{k}^{\rho}$ determined by the probe field wave vector to read
\begin{eqnarray}
 \mathbf{G}_{\text{1D}}(z_i,z_j,\omega_p,\bm{k}^{\rho})&\simeq&\frac{ i}{2 k_z}  \left[\bm{p}_\nu\left(\text{z}_i\right)\bm{q}_\nu\left(\text{z}_j\right)\Theta\left(\text{z}_i-\text{z}_{j}\right) \right. \nonumber \\
&+&   \left. \bm{p}_\nu\left(\text{z}_j\right)  \bm{q}_\nu\left(\text{z}_i\right)\Theta \left(\text{z}_j-\text{z}_{i}\right) \right] \, .
\end{eqnarray}
Here, $k_z$ is the $z$-component of the wave number, the quantities $\bm{p}_\nu$ and $\bm{q}_\nu$  represent the fields produced in the cavity by a grazing incidence x-ray pulse of unit strength incident upon the cavity from its lower and upper sides, respectively,  and $\Theta(z)$ denotes the Heaviside step function.

The spin-exchange and decay rates defined in Eqs.~(\ref{c1}) further depend on the dipole matrix element $\mathbf{d}$. This quantity (in the case of $^{57}\mathrm{Fe}$ the magnetic dipole matrix element) can be connected to the radiative decay rate of a single nucleus $\Gamma_r$ as shown for instance in Ref.~\cite{kong2014field}.  In turn, the radiative decay rate can be written with the help of the spontaneous decay rate of the nuclear excited state of a single nucleus $\Gamma_0$ taking into account the internal conversion channel and the internal conversion coefficient $\alpha$ as  $\Gamma_0=(1+\alpha)\Gamma_r$.
For the 14.4 keV transition in  $^{57}\mathrm{Fe}$, the internal conversion coefficient is approximately 8.6.

The total Green function for the system includes apart from the cavity Green function also an additional vacuum contribution,  $\mathbf{G}_{\mathrm{tot}}\left(\mathbf{r}_{i},\mathbf{r}_{j},\omega_p \right)=\mathbf{G}\left(\mathbf{r}_{i},\mathbf{r}_{j},\omega_p \right)+\mathbf{G}_{\text{vac}}\left(\mathbf{r}_{i},\mathbf{r}_{j},\omega_p \right)$. We further assume that the nucleus-nucleus couplings via the cavity channel play a significant role and $\mathbf{G}_{\text{vac}}\left(\mathbf{r}_{i},\mathbf{r}_{j},\omega_p \right)$ only provides the spontaneous decay of a single nucleus $\Gamma_0$, whose value we take from experiments. In the following, we proceed to apply our approach  for some specific cases of thin-film cavities.

\subsection{A single nuclear layer \label{1layer}}

We first consider the case of a single nuclear layer embedded in the x-ray cavity at $z=z_0$, as illustrated in  Fig.~\ref{system}.  The cavity consists of a sandwich  of Pd and C layers with one embedded $^{57}\mathrm{Fe}$ layer. Since the nuclear layer is very thin compared with the wavelength of the standing wave in the cavity under the grazing incidence with a few mrad, we assume that the electric field for all the nuclei in the thin layer is the same, i.e., $\mathbf{G}_{\text{1D}}(z_i,z_j,\omega_p,\bm{k}^{\rho})\simeq \mathbf{G}_{\text{1D}}(z_0,z_0,\omega_p,\bm{k}^{\rho})$. In order to get rid of the dependence in the transversal plane, we define similarly to Refs.~\cite{masson2019atomicwaveguide,Lentrodt2020} the collective nuclear spin-wave operators for the nuclei in the layer,
\begin{equation}
 \hat{S}(\bm{k}^{\rho})=\frac{1}{\sqrt{N}}\sum_{i=1}^{N}e^{-i\bm{k}^{\rho}\bm{\rho}_i}\hat{\sigma}_{ge}^{i}\, .
 \label{s1}
\end{equation}
Under the assumption of translational invariance, the spin-wave operators diagonalize the Hamiltonian. Following Ref.~\cite{Lentrodt2020}, we can perform the change of basis and the integration over $\bm{k}^{\rho}$ appearing in Eq.~\eqref{G2}. Due to the translational symmetry and in the low-saturation regime, only the subspace with wave vector $\bm{k}^{\rho}_p$, the transversal component of the incident wave vector $\bm{k}_p$, is driven by the probe field. 
Restricted to this subspace, the Hamiltonian and Lindblad operators are simplified and read
\begin{eqnarray}
  \hat{H}&=&-\hbar\Delta \hat{S}^{\dagger}(\bm{k}^{\rho}_p)\hat{S}(\bm{k}^{\rho}_p)+\hbar Ng\hat{S}^{\dagger}(\bm{k}^{\rho}_p)\hat{S}(\bm{k}^{\rho}_p)
\nonumber \\ 
 &-&\hbar{\sqrt{N}}\left[\Omega\hat{S}^{\dagger}(\bm{k}^{\rho}_p)+\Omega^*\hat{S}(\bm{k}^{\rho}_p)\right]\, ,
\end{eqnarray}
and 
\begin{eqnarray}
 L[\hat{\rho}]&=&-\frac{N\gamma+\Gamma_0}{2}\left[\hat{S}^{\dagger}(\bm{k}^{\rho}_p)\hat{S}(\bm{k}^{\rho}_p)\hat{\rho}+\hat{\rho}\hat{S}^{\dagger}(\bm{k}^{\rho}_p)\hat{S}(\bm{k}^{\rho}_p) \right. \nonumber \\
 &-& \left. 2\hat{S}(\bm{k}^{\rho}_p)\hat{\rho}\hat{S}^\dagger(\bm{k}^{\rho}_p)\right]\, ,
\end{eqnarray}
where  $\Omega=\mathbf{d}^{*}\cdot\hat{\mathbf{E}}_{\text{1D}}^{+}(z_{0})/\hbar$ is the Rabi frequency of a single nucleus. Using the thin-layer approximation $\mathbf{G}_{\text{1D}}(z_i,z_j,\omega_p,\bm{k}^{\rho})\simeq \mathbf{G}_{\text{1D}}(z_0,z_0,\omega_p,\bm{k}^{\rho})$, the spin-exchange and decay rates no longer depend on the nuclear indices $i,j$ and can be written in simplified form
\begin{align}
	g&=\frac{\mu_{0}\omega_{p}^{2}}{\hbar A}\, \mathbf{d}^*\cdot \text{Re}\left[\mathbf{G}_{\text{1D}}(z_0,z_0,\omega_p,\bm{k}^{\rho}_p)\right]\cdot \mathbf{d}\, ,  \nonumber \\
	\gamma&=\frac{2\mu_{0}\omega_{p}^{2}}{\hbar A}\, \mathbf{d}^*\cdot \text{Im}\left[\mathbf{G}_{\text{1D}}(z_0,z_0,\omega_p,\bm{k}^{\rho}_p)\right]\cdot\mathbf{d}\, , 
	\label{gamma}
\end{align}
where $A$ is the transversal area denoted in Ref.~\cite{Lentrodt2020} as parallel quantization area. The area factor is not determined analytically but is contained in the fitting procedure  described in Section \ref{numres}. 


 We work in the Heisenberg representation. Using the master equation for the density matrix, the expectation value of the collective  coherence $\left(S=\langle\hat{S}(\bm{k}^{\rho}_p)\rangle\right)$ governed by the Heisenberg equations reads 

\begin{equation}
 \dot{S}=i\left(\Delta+i\frac{\Gamma_0}{2}\right)S+i{\sqrt{N}}\Omega+ i\left(Ng+i\frac{N\gamma}{2}\right)S\, .
\end{equation}
The coherences will evolve towards a steady state described by the  solution of the equation  $\dot{S}=0$. We obtain
\begin{equation}
 S=-\frac{\sqrt{N}\Omega}{\Delta+Ng+i\frac{N\gamma+\Gamma_0}{2}}\, .
 \label{S1layer}
\end{equation}
From the equation above we can derive a simple interpretation for the behaviour of the collective resonant scattering of the nuclei in the embedded $^{57}\mathrm{Fe}$ layer. We observe that in the denominator the real part is shifted by $Ng$, while the imaginary part is increased by $N\gamma$. Thus, the thin nuclear layer acts like a giant ``macro-nucleus'' with a collective frequency shift $Ng$ (the collective Lamb shift) and a superradiant decay rate $N\gamma$.

\subsection{Multi-layers \label{multi-theo}}
We now proceed to a more complicated case with $n_0$ nuclear layers embedded in the cavity. 
 We start again from the general Hamiltonian in Eq.~(\ref{h1}). All nuclei in the same layer have the same $z$ coordinate  and we denote with $z_{l}$ and $N_l$ for $l=1,2...n_0$  the position of and number of nuclei in each layer, respectively. Using the approximations introduced in Subsection~\ref{1layer}, the spin-exchange and decay rates defined in Eq.~(\ref{c1}) are the same for all nuclei from the same layer but different for the nuclei from different layers. We note here that this approximation might not always hold in practice, as the standing wave might not have a much larger wavelength than the nuclear layer thickness. This case will be discussed later on in a practical example presented in Section~\ref{numres}.

Defining  the  collective nuclear spin operator $S_l$ for each layer $l$ according to Eq.~(\ref{s1}), we derive the Hamiltonian for the multi-layer system { 

\begin{align}
	\hat{H}=&-\hbar\Delta\sum_{l=1}^{n_0}\hat{S}_{l}^{\dagger}(\bm{k}^{\rho}_p)\hat{S}_{l}(\bm{k}^{\rho}_p) \nonumber \\
	& -\hbar\sum_{l,m=1}^{n_0}J_{lm}\hat{S}_{l}^{\dagger}(\bm{k}^{\rho}_p)\hat{S}_{m}(\bm{k}^{\rho}_p)  \nonumber \\
	&-\hbar\sum_{l=1}^{n_0}\left[\Omega_{l}\hat{S}_{l}^{\dagger}(\bm{k}^{\rho}_p)+\Omega_{l}^*\hat{S}_{l}(\bm{k}^{\rho}_p)\right]\, , 
	\label{h2}
\end{align}
and the Lindblad operators  
\begin{align}
	L[\hat{\rho}]=&-\sum_{l,m=1}^{n_0}\frac{\Gamma_{lm}}{2}\left[\hat{S}_{l}^{\dagger}(\bm{k}^{\rho}_p)\hat{S}_{m}(\bm{k}^{\rho}_p)\hat{\rho} \nonumber \right.\\
	& \left. +\hat{\rho}\hat{S}_{l}^{\dagger}(\bm{k}^{\rho}_p)\hat{S}_{m}(\bm{k}^{\rho}_p)-2\hat{S}_{l}(\bm{k}^{\rho}_p)\hat{\rho}\hat{S}_{m}^{\dagger}(\bm{k}^{\rho}_p)\right] \nonumber \\
	&-\frac{\Gamma_{0}}{2}\sum_{l=1}^{n0}\left[\hat{S}_{l}^{\dagger}(\bm{k}^{\rho}_p)\hat{S}_{l}(\bm{k}^{\rho}_p)\hat{\rho}+\hat{\rho}\hat{S}_{l}^{\dagger}(\bm{k}^{\rho}_p)\hat{S}_{l}(\bm{k}^{\rho}_p) \right. \nonumber \\
	&\left. -2\hat{S}_{l}(\bm{k}^{\rho}_p)\hat{\rho}\hat{S}_{l}^{\dagger}(\bm{k}^{\rho}_p)\right]\, ,
	\label{l2}
\end{align}
where 
\begin{align}
 J_{lm}&=\sqrt{N_{l}N_{m}}\, \frac{\mu_{0}\omega_{p}^{2}}{\hbar A}\,\mathbf{d}^*\cdot \text{Re}\left[\mathbf{G}_{\text{1D}}(z_l,z_m,\omega_p,\bm{k}^{\rho}_p)\right]\cdot \mathbf{d}\, , \nonumber \\
\Gamma_{lm}&=\sqrt{N_{l}N_{m}}\,\frac{2\mu_{0}\omega_{p}^{2}}{\hbar A}\,\mathbf{d}^*\cdot\text{Im}\left[\mathbf{G}_{\text{1D}}(z_l,z_m,\omega_p,\bm{k}^{\rho}_p)\right]\cdot\mathbf{d}\, , \nonumber \\
\Omega_{l}&=\sqrt{N_l}\, \mathbf{d}^{*}\cdot\hat{\mathbf{E}}_{\text{1D}}^{+}(z_{l})/\hbar \, .
\label{p2}
\end{align}
Here, the indices $l,m$ indicate the layers.  
Recalling the interpretation introduced at the end of Subsection~\ref{1layer} for a single-layer cavity, we can regard $J_{lm}$ (the real part of the inter-layer coupling) as a coherent coupling  or spin-exchange rate betweeen ``macro-nuclei'' and 
$\Gamma_{lm}$ (the imaginary part of the inter-layer coupling) as an incoherent coupling or decay rate.

The dynamics of the nuclear coherences are driven by the Heisenberg equations 

\begin{equation}
\dot{S}_{l}=i\left(\Delta+i\frac{\Gamma_0}{2}\right)S_{l}+i\Omega_{l}+i\sum_{m=1}^{n_0}G_{lm}S_{m}\, .
\end{equation}
where 
\begin{align}
G_{lm}&=J_{lm}+i\Gamma_{lm}/2 \nonumber \\
&=\sqrt{N_{l}N_{m}}\frac{\mu_{0}\omega_{p}^{2}}{\hbar A}\,\mathbf{d}^*\cdot\mathbf{G}_{\text{1D}}(z_l,z_m,\omega_p,\bm{k}^{\rho}_p)\cdot \mathbf{d}
\end{align}
is determined by the Green function.

For the steady state condition $\dot{S}=0$ we obtain $ \vec{S}=-\mathbb{M}^{-1}\vec{\Omega}$ with
\begin{equation}
  \ \mathbb{M}=\left(\Delta+i\frac{\Gamma_0}{2}\right)\mathbb{1}+\emph{G}\, .
\end{equation}
Here $\vec{S}=\left(S_{1},\cdots S_{n_0}\right)$ and $\Omega=\left(\Omega_1,\cdots\Omega_{n_0}\right)$ are $n_0$-dimensional vectors, and $\mathbb{M}$ is a $n_0\times n_0$ matrix which is determined by $\Delta$, $\Gamma_0$ and the matrix $\emph{G}$ consisting of elements $G_{ij}$.

\subsection{Input-output formalism}

So far, our model provides the expectation value of the nuclear coherences. However, the observable in experiments is the energy- or incidence angle-dependent cavity reflectivity.
The connection is provided by simple expressions that connect the field at the edge of the cavity to the nuclear coherences. The output operator for the reflectivity spectrum is defined as $\hat{a}_{\text{out}}=\hat{\mathbf{E}}^{+}_{\text{out}}(z_{\text{top}}) e^{ i\bm{k}^{\rho}_p\bm{\rho}}  $ where $z_{\text{top}}$ is the position of the incidence boundary. The reflectance is written as
\begin{equation}
R=\frac{\left<\hat{a}_{\text{out}}\right>}{a_{\text{in}}}\, ,
\end{equation}
where $a_{\text{in}}$ is the input field.  The field at any point in space can be reconstructed in terms of the coherences. For the one-layer case, the expression for the field operator is given by \cite{asenjo2017atom} 
\begin{equation}
 \hat{\mathbf{E}}^{+}_{\text{out}}(z)=\hat{\mathbf{E}}^{+}_{\text{1D}}(z)+\frac{\mu_{0}\omega_{p}^{2}\sqrt{N}}{A}\mathbf{G}_{\text{1D}}(z,z_0,\omega_p,\bm{k}^{\rho}_p)\cdot \mathbf{d}\, \hat{S}(\bm{k}^{\rho}_p)\, ,
 \label{x}
\end{equation}
where  $\hat{\mathbf{E}}^{+}_{\text{1D}}(z)$ stands for the field scattered by the cavity in the absence of the resonant nuclei and the second term can be considered as the field rescattered by the nuclei. 
For the multi-layer case, the field operator reads \cite{asenjo2017atom}
\begin{eqnarray}
 &&\hat{\mathbf{E}}^{+}_{\text{out}}(z)=\hat{\mathbf{E}}^{+}_{\text{1D}}(z)\nonumber \\
 &&+\frac{\mu_{0}\omega_{p}^{2}}{A}\sum_{l=1}^{n_0}\sqrt{N_l}\mathbf{G}_{\text{1D}}(z,z_l,\omega_p,\bm{k}^{\rho}_p)\cdot\mathbf{d}\, \hat{S}_{l}(\bm{k}^{\rho}_p)\, .
 \label{y}
\end{eqnarray}

Another interesting observable which can easily be calculated with our formalism  is the photon correlation function. For instance, following Ref.~\cite{PhysRevA.88.043828} the second order correlation function of x-ray photons over a time interval $\tau$ is accessed as
\begin{equation}
	g_{2}\left(\tau\right)=\frac{\left<\hat{a}_{\text{out}}^\dagger\left(0\right)\hat{a}_{\text{out}}^\dagger\left(\tau\right)\hat{a}_{\text{out}}\left(\tau\right)\hat{a}_{\text{out}}\left(0\right)\right>}{\left<\hat{a}_{\text{out}}^\dagger\left(0\right)\hat{a}_{\text{out}}\left(0\right)\right>^2}\, .
\end{equation}
We recall here that although omitted in the notations, all operators are time-dependent in the Heisenberg picture used here.
The second order correlation function can be used to investigate the x-ray photon statistics, such as photon bunching and antibunching which can not be calculated by the semiclassical  Parratt formalism or the layer-formalism.

\section{Numerical Results}
\label{numres}
We now proceed to apply the formalism described above to  several thin-film cavity structures with one, two or thirty embedded $^{57}\mathrm{Fe}$ layers. We validate our approach by carefully comparing our predictions for the cavity reflectivity with simulations with CONUSS \cite{sturhahn2000conuss}  for different layer structures. Particular features which are well reproduced by the model are highlighted  and physical interpretations are presented. 

 Our approach is not ab initio, as we do require one fit parameter. The latter is the ratio of two factors: the unknown number of nuclei $N$, and the transversal area factor $A$  entering the spin-exchange and decay rates in Eqs.~\eqref{gamma} and \ref{p2}. The reflectivity observable based on the field expressions \eqref{x}  and \eqref{y} depends on the term $\sqrt{N} \mathbf{G}_{\text{1D}}(z,z_0,\omega_p,\bm{k}^{\rho}_p)\cdot \mathbf{d}\, S/A$, which in turn depends on the area density factor $N/A$.  We note that 
Ref.~\cite{Lentrodt2020} which presents an ab initio Green function model for thin-film x-ray cavity interprets the area factor $A$ as a parallel  quantization area 
 and determines the planar nuclear density from the sample nuclear density.

\subsection{Cavities with a single embedded nuclear layer}
We consider a layer structure as the one reported in  Ref.~\cite{rohlsberger2010collective}, namely (2.2 nm Pt)/(16 nm C)/(0.6 nm $^{57}\text{SS}$)/(16 nm C)/(13 nm Pt)  where $^{57}\text{SS}$ is stainless steel containing 
$^{57}\mathrm{Fe}$-enriched iron (95\%). For such a cavity, our model predicts a reflectivity with the expression
\begin{equation}
 R=R_0+i\frac{C}{\Delta+Ng+i\left(N\gamma+\Gamma_0\right)/2} \, ,
 \label{r1}
\end{equation}
where $C$ is a constant depending on the incidence angle $\varphi$  and $N/A$, and $R_0$ is the reflectivity from the bare cavity without considering the interaction with the nuclei.  The factor $N/A$ is the only fit parameter of our model and is a function of the resonant nuclear layer thickness and the density of $^{57}\mathrm{Fe}$ nuclei in the layer. The latter in turn depends on the degree of $^{57}\mathrm{Fe}$-enrichment and the chemical composition. Practically, we can obtain $N/A$ as a scaling parameter by fitting once a reflectivity spectrum with the corresponding CONUSS predictions. For the same nuclear layer thickness and  composition, the obtained value $N/A$ can be used for any cavity structure and for all incidence angles. Furthermore, for the same composition, $N/A$ scales linearly with the layer thickness,  as expected for a planar nuclear density.

Our numerical results are compared with predictions by CONUSS in Fig.~\ref{fig1} for the resonant angle of the first guided mode at $\varphi_0=2.464\ \text{mrad}$. The figure shows the two main features predicted by Eqs.~(\ref{S1layer}) and (\ref{r1}): a frequency shift from the nuclear transition (corresponding to $\Delta=0$ in the plot), and a line broadening compared to spontaneous decay as known from single nuclei. In the literature these features are known as the collective Lamb shift \cite{rohlsberger2010collective} and superradiant decay \cite{NFS-Bible}. Our predictions are in excellent agreement with the numerical CONUSS simulations (green dashed line).  At the cavity resonant incidence angles, the scattered field due to the bare cavity in the absence of nuclei [$\hat{\mathbf{E}}^{+}_{\text{1D}}(z)$ in Eq.~(\ref{x})] is suppressed, $R_0$ is zero and  
the superradiant decay determines the width of the Lorentz shape in Fig.~\ref{fig1}.  This situation changes for deviating incidence angles, where the non-resonant cavity reflectivity $R_0$ becomes non-zero and the total reflectivity is no longer a Lorentz profile.

\begin{figure}
\includegraphics[width=0.3\textwidth]{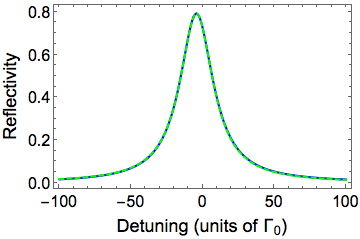}
\caption{\label{fig1} Reflectivity of a cavity with a single $^{57}\mathrm{SS}$ layer calculated for the resonant angle of the first guided mode at $\varphi=2.464\ \text{mrad}$ with the Green function formalism (blue solid line) and with CONUSS (green dashed line). See text for further explanations.}
\end{figure} 

We now investigate in more detail the behaviour of the superradiant decay as a function of the exact placement of the nuclear layer in the cavity. We consider the  cavity structure   Pt(2 nm)/C(40 nm)/Pt(10 nm) with a 1 nm $^{57}\mathrm{Fe}$ layer placed in the cavity at position $z_0$ measured starting from the top. We calculate reflectivity spectra for incidence angles around the corresponding resonant angle of the third guided mode for  $z_0=$2.5 nm, 7.5 nm, 12 nm and 16.5 nm. A comparison between our result and CONUSS simulations for the example of $z_0$=12 nm is shown in Fig.~\ref{figgc}.  Also in this case the agreement is excellent. We have checked that the same holds for the other three cases with $z_0=$2.5 nm, 7.5 nm  and 16.5 nm, not presented here.

\begin{figure}[]
	\centering
	
	\subfigure{
		\label{figgca} 
		\includegraphics[width=2.2in]{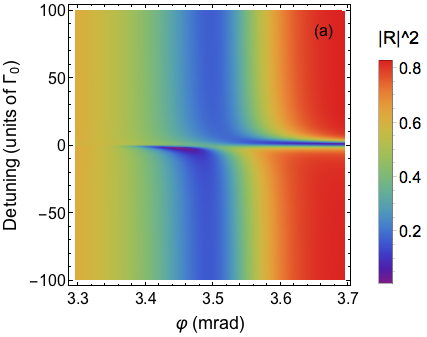}}
	\hspace{1in}
	\subfigure{
		\label{figgcb} 
		\includegraphics[width=2.2in]{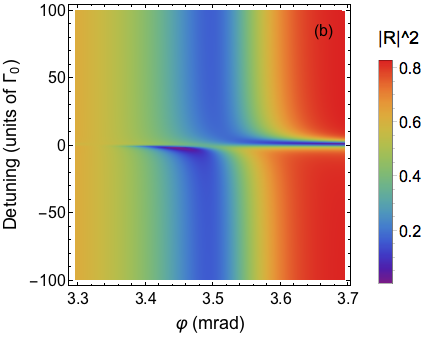}}
	\caption{Calculated energy- and angle-dependent reflectivity 
	for the  Pt(2 nm)/C(40 nm)/Pt(10 nm) cavity with the 1 nm $^{57}\mathrm{SS}$ layer placed at $z_0$=12 nm. (a) Numerical data from the Green function model. (b) Numerical results from CONUSS. See text for further explanations.
	}
	\label{figgc} 
\end{figure}


An interesting aspect when varying the position of the resonant layer inside the cavity is the shape of the superradiant decay $N\gamma$ as a function of the incidence angle. In a single-mode Jaynes-Cummings model,  the collectively enhanced decay rate  $\Gamma_C$ in the cavity as a function of the incidence angle is given by a Lorentz profile
 \cite{PhysRevA.88.043828}
\begin{equation}
\Gamma_C=\frac{2|\tilde{g}|^2\kappa}{\kappa^2+\Delta_C^2}\, ,
\label{f2}
\end{equation}
where $\tilde{g}$ is the coupling between the nuclei and the cavity, $\kappa$ is the cavity decay  and $\Delta_C=\omega_0-\omega_c$ is the cavity detuning, proportional to the deviation from the incidence angle $\Delta\varphi$. 

In our model  $N\gamma$ can be calculated using the second of Eqs.~\ref{gamma}. In Fig.~\ref{fano} we present the  ratio $F_p=N\gamma/\Gamma_0$ for the four considered $z_0$ values as a function of the angular detuning $\Delta \varphi$ around the first minimum of the reflectivity. Surprisingly, the Lorentz profile appears to describe only the case of the nuclear layer placed in the antinode of the guided mode standing wave.  For the other positions, the superradiant decay displays a Fano instead of a Lorentzian shape as a function of the incidence angle.   The dependence of the Fano profile on $z_0$ can be quantified by fitting the calculated $F_p$
with the function $a\frac{|q+b\Delta\varphi|^2}{1+b^2{\Delta\varphi}^2}$  using $a,q,b,\varphi_C$ as fitting parameters. The results are  presented in Fig.~\ref{fano}. When the nuclear layer is placed at the antinode of the standing wave in the cavity ($z_0=7.5$ nm), the fitting parameter $|q|=150.8$ is very large [see Fig.~\ref{fano}(b)] and the line shape closely resembles a Lorentz line, being consistent with the Jaynes-Cummings expression (\ref{f2}).  If the nuclear layer is not at the antinode,  $F_p$ is asymmetric and it can be fitted by a Fano line shape with Fano asymmetry parameters $|q|=3.5$, 2.9 and 1.9 as shown in Figs.~\ref{fano}(a), (c) and (d), respectively. This proves the strength of our model which in contrast to the single-mode Jaynes-Cummings model can handle in its description all cavity modes.

\begin{figure}{}
	\includegraphics[width=0.45\textwidth]{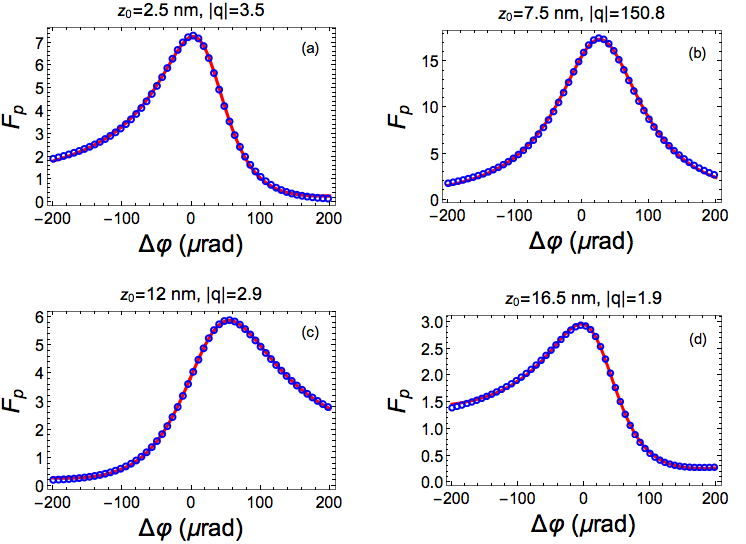}
	\caption{\label{fano} The factor $F_p$ as a function of the deviation angle $\Delta\varphi$ for different positions $z_0$ of the nuclear layer  $^{57}\mathrm{Fe}$. Our theoretical values (blue circles) are fitted by a Fano profile (red line) as described in the text. The obtained Fano asymmetry parameter $q$ is given above each graph.  }
\end{figure} 

\subsection{Results for multi-layer systems}

We now check the accuracy of our formalism for more complex examples with more than one nuclear layer placed in the cavity. Few experiments have already been performed for such cavities, for instance, the first demonstration of electromagnetically induced transparency (EIT) in the x-ray regime \cite{rohlsberger2012electromagnetically} and of the collective strong coupling of x-rays \cite{haber2016,haber2017rabi}. We first consider the two cavity structures investigated in Ref.~\cite{rohlsberger2012electromagnetically}. These two cavities contain two embedded $^{57}\mathrm{Fe}$ layers and differ in the exact placement of the latter. Both cavities consist of a 
Pt(3 nm)/C(38 nm)/Pt(10 nm) sandwich structure, each containing two $^{57}\mathrm{Fe}$ layers, placed  one at the node and one at the antinode of the cavity field.  Ref.~\cite{rohlsberger2012electromagnetically} considers 2-nm and 3-nm thick layers. At this thickness,   the
iron  layers order ferromagnetically with the magnetization confined to
the plane of the films. The magnetic hyperfine interaction lifts the
degeneracy of the nuclear magnetic sublevels, leading to four allowed
magnetic dipole transitions for the given scattering geometry, where the magnetization is aligned parallel to the wavevector of the incident photons. As the generated intrinsic magnetic field has 33 T, the driven transitions do not overlap and can be considered separately.  Here we perform simulations for 2-nm thick 
$^{57}\mathrm{Fe}$ layers in order to reproduce Fig.~1 of Ref.~\cite{rohlsberger2012electromagnetically}.

The structure for which EIT features appear for each of the four hyperfine transitions has the first nuclear layer at the node centered at  $z_0=15.2$ nm and the second one at the antinode of the standing wave at approximately  $z_0=24.2$ nm. At zero detuning, the reflectivity presents a clear dip, which resembles transparency and is therefore attributed to  EIT-like effects. In the second structure, the positions of the nuclear layers are inverted such that the first layer is in an antinode at approximately $z_0=24.2$ nm and the second layer in the following node at  $z_0=33.2$ nm.  For this cavity structure, the EIT feature, i.e., the reflectivity dip in the scattered spectrum, disappears. The calculated reflectivity spectra for both structures are shown in Fig.~\ref{fig4} for the resonant angles of the third guiding modes which around  3.57 mrad. The comparison with CONUSS shows excellent agreement and provides 
  a strong evidence for the validity of our method.

\begin{figure}{}
	\includegraphics[width=0.48\textwidth]{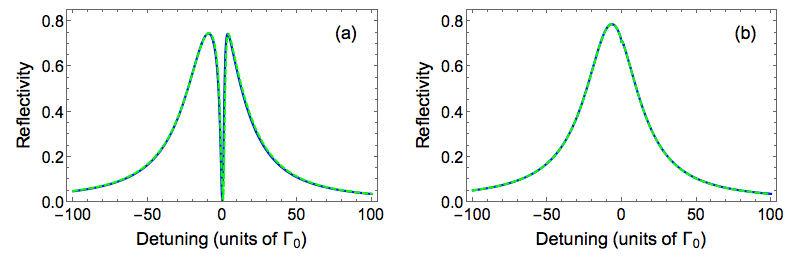}
	\caption{\label{fig4} Energy-dependent reflectivity spectrum for the  two-layer structures calculated for the resonant angle of the third guiding mode with the Green function formalism (blue solid line) and CONUSS (green dashed line).  (a) Numerical data for the node-antinode structure. A dip occurs around the resonant energy. (b) Numerical results for the antinode-node structure. A single resonant line is obtained and the EIT dip disappears. See text for further explanations.}
\end{figure} 

The physical picture of the two-layer structure emerging from our model expressions is easy to follow. Each resonant layer has an individual collective Lamb frequency shift $J_{ii}$ and superradiance decay rate $\Gamma_{ii}$ where $i=1,2$. Moreover, a complex interlayer coupling $J_{12}+i\Gamma_{12}/2$ comes into play. The absolute value of this complex coupling is not negligible at the resonant angles, being on the order of few $\Gamma_{0}$. 
For both cavity structures, the complex eigenstates of the system can be depicted as one broad and one narrow Lorentzian. The difference between the two cavity systems is that for the EIT case, the two eigenstates interfere and a dip appears 
around the resonant energy as presented in Fig.~\ref{fig4}(a). For the other cavity structure, the narrow eigenstate almost vanishes, i.e., it becomes a dark state which is not probed by the x-ray pulse.  In the absence of interference, we observe only the broad eigenstate as shown in Fig.~\ref{fig4}(b). We note here that this physical picture is slightly different from the one presented in Ref.~\cite{rohlsberger2012electromagnetically}, where it is argued that the coupling strength between the two layers plays the role of EIT control field. For the antinode-node structure, Ref.~\cite{rohlsberger2012electromagnetically} attributes the disappearance of the transparency dip to the very small value of the coupling. However, we find that the absolute value of the complex coupling $J_{12}+i\Gamma_{12}/2$ is on the same order of magnitude for the two cases,   3.8$\Gamma_{0}$ for the EIT result in Fig.~\ref{fig4}(a)  and  3.2$\Gamma_{0}$ for the system in Fig.~\ref{fig4}(b).  Our model confirms the similar  conclusion reached in Ref.~\cite{heeg2015collective} on the basis of the previously available quantum model for x-ray thin film cavities.

In the last part of this section, we test our model for the multilayer structure with thirty nuclear layers considered in Ref.~\cite{haber2016}. All numerical simulations presented in Ref.~\cite{haber2016} are calculated from semiclassical methods (the transfer matrix method similar to the classical Parratt algorithm and CONUSS) and to the best of our knowledge, so far no quantum model has been directly applied to this case because of the structure complexity with a large number of  nuclear layers. The multilayer sample consists of 30 bilayers of (1.12 nm $^{57}\mathrm{Fe}$/1.64 nm $^{56}\mathrm{Fe}$), which are probed with x-rays in incidence angles between $15$ mrad and $17$ mrad \cite{haber2016}.

We find that in this case, the wavelength of the standing wave is only 3 to 4 times larger than the layer thickness, such that our approximation that all nuclei within one layer feel the same cavity field is no longer accurate. In order to tackle this problem, we slice each nuclear layer in four sub-layers  each of thickness 0.28 nm, and approximate for each sub-layer that all nuclei are at the same position in coordinate $z$. Thus, each nuclear layer consists of four ``macro-nuclei'' instead of one, which we consider according to the procedure described in Section \ref{multi-theo}.
 Our numerical simulations for the reflectivity as a function of angular and frequency detuning are compared with CONUSS numerical data in Fig.~\ref{fig8}. The displayed agreement is also for this complex case excellent. We note that the picture changes dramatically if we would not adjust our procedure to accommodate the large layer thickness. Considering all nuclei in each layer to experience the same cavity field (without further separation in sub-layers), the reflectivity changes and the splitting around $\varphi=16$ mrad disappears. 

\begin{figure}[]
	\centering
	
	\subfigure{
		\label{fig6a} 
		\includegraphics[width=2.2in]{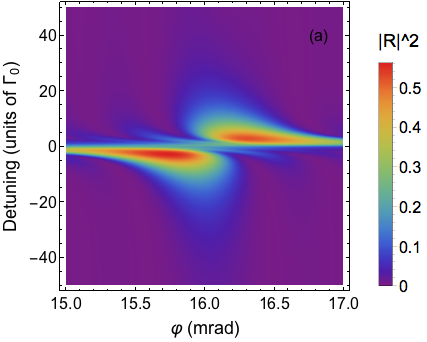}}
	\hspace{1in}
	\subfigure{
		\label{fig6b} 
		\includegraphics[width=2.2in]{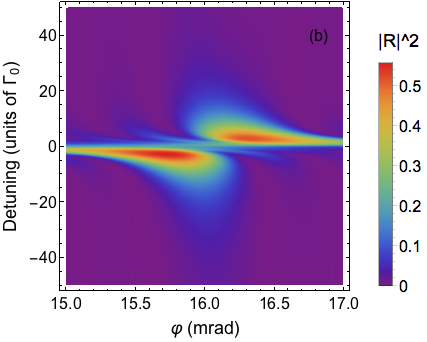}}
	\caption{Calculated energy- and angle-dependent reflectivity 
	for the multilayer structure in  Ref.~\cite{haber2016}. (a) Numerical data from the Green function model. (b) Numerical results from CONUSS. See text for further explanations.
	}
	\label{fig8} 
\end{figure}


\section{Conclusion}
\label{4}

This paper adapts a Green function formalism known from superradiant systems in quantum optics of atoms to x-ray thin-film cavities with embedded nuclear layers.  An important approximation which significantly simplifies the calculations is based on the small thickness of the nuclear layer(s) as opposed to the field cavity wavelength. The advantages of the formalism are its versatility and the fact that it only requires one fit parameter, which solely depends on the nuclear layer thickness and composition. Another advantage is 
its ability to predict also quantum properties of the scattered field such as higher order correlation functions,  which is not available in semi-classical models.  Because there is no restriction to the validity based upon excitation number, the formalism in principle provides a route toward modeling multi-photon quantum effects, as might be achievable with intense XFEL light. We have benchmarked the model against observables calculated with semi-classical methods based on the layer formalism implemented in the computer package CONUSS \cite{sturhahn2000conuss}. The simulations show excellent agreement for thin-film cavity structures with one, two or thirty embedded nuclear layers. The model provides clear intuitive pictures of the underlying physics and correctly reproduces features that go beyond the single-mode Jaynes-Cummings model.
 We believe that this formalism can be used as a versatile tool for the calculation of scattering spectra of thin-film cavities of any structure, and eventually help the future design of x-ray photonic devices.

\begin{acknowledgments}
 We thank T. Shi and Y. Chang for helpful discussions. This work is part of and supported by the DFG
Collaborative Research Center ``SFB 1225 (ISOQUANT)''. XJK acknowledges  support by the National Natural Science Foundation of China (NSFC) under Grant No. 11904404. DEC acknowledges support from MINECO Severo Ochoa Grant No. CEX2019-000910-S, CERCA Programme/Generalitat de Catalunya, Fundació Privada Cellex, Fundació Mir-Puig, the European Research Council (ERC) under the European Union’s Horizon 2020 research and innovation programme (Grant Agreement No. 639643), and Plan Nacional Grant ALIQS (funded by MCIU, AEI, and FEDER).

\end{acknowledgments}

\bibliography{green}

\end{document}